\documentclass[%
 aip,
 amsmath,amssymb,
 reprint,%
]{revtex4-1}

\usepackage{graphicx}
\usepackage{dcolumn}
\usepackage{bm}
\usepackage{fixltx2e}
\usepackage[utf8]{inputenc}
\usepackage[T1]{fontenc}
\usepackage{mathptmx}
\usepackage{graphicx}
\usepackage{float}
\usepackage{color}
\usepackage[normalem]{ulem} 

\begin{document}

\preprint{AIP/123-QED}

\title[The anisotropic quasi-static permittivity of single-crystal $\beta$-Ga$_2$O$_3$]{The anisotropic quasi-static permittivity of single-crystal $\beta$-Ga$_2$O$_3$}

\author{Prashanth Gopalan}
\affiliation{Department of Electrical and Computer Engineering, The University of Utah, Salt Lake City, UT 84112}
\email{berardi.sensale@utah.edu}
\author{Sean Knight}
\affiliation{Terahertz Materials Analysis Center and Competence Center for III-Nitride Technology C3NiT - Janz\'{e}n, Department of Physics, Chemistry and Biology (IFM), Link\"{o}ping University, Link\"{o}ping, SE 58183, Sweden}
\author{Ashish Chanana}
\affiliation{Department of Electrical and Computer Engineering, The University of Utah, Salt Lake City, UT 84112}
\author{Megan Stokey}
\affiliation{Department of Electrical and Computer Engineering, University of Nebraska-Lincoln, Lincoln, NE 68588, USA}
\author{Praneeth Ranga}
\affiliation{Department of Electrical and Computer Engineering, The University of Utah, Salt Lake City, UT 84112}
\author{Michael A.Scarpulla}
\affiliation{Department of Electrical and Computer Engineering, The University of Utah, Salt Lake City, UT 84112}
\affiliation{Department of Material Science and Engineering, The University of Utah, Salt Lake City, UT 84112}
\author{Sriram Krishnamoorthy}
\affiliation{Department of Electrical and Computer Engineering, The University of Utah, Salt Lake City, UT 84112}
\author{V. Darakchieva}
\affiliation{Terahertz Materials Analysis Center and Competence Center for III-Nitride Technology C3NiT - Janz\'{e}n, Department of Physics, Chemistry and Biology (IFM), Link\"{o}ping University, Link\"{o}ping, SE 58183, Sweden}
\author{Zbigniew Galazka}
\affiliation{Leibniz-Institut f\"{u}r Kristallz\"{u}chtung, Berlin, 12489, Germany}
\author{Klaus Irmscher}
\affiliation{Leibniz-Institut f\"{u}r Kristallz\"{u}chtung, Berlin, 12489, Germany}
\author{Andreas Fiedler}
\affiliation{Leibniz-Institut f\"{u}r Kristallz\"{u}chtung, Berlin, 12489, Germany}
\author{Steve Blair}
\affiliation{Department of Electrical and Computer Engineering, The University of Utah, Salt Lake City, UT 84112}
\author{Mathias Schubert}
\affiliation{Department of Electrical and Computer Engineering, University of Nebraska-Lincoln, Lincoln, NE 68588, USA}
\affiliation{Terahertz Materials Analysis Center and Competence Center for III-Nitride Technology C3NiT - Janz\'{e}n, Department of Physics, Chemistry and Biology (IFM), Link\"{o}ping University, Link\"{o}ping, SE 58183, Sweden}
\affiliation{Leibniz Institut f\"{u}r Polymerforschung e.V., 01069 Dresden, Germany}
\author{Berardi Sensale-Rodriguez}
\affiliation{Department of Electrical and Computer Engineering, The University of Utah, Salt Lake City, UT 84112}

\date{\today}

\begin{abstract}
The quasi-static anisotropic permittivity parameters of electrically insulating gallium oxide ($\beta$-Ga$_2$O$_3$) were determined by terahertz spectroscopy. Polarization-resolved frequency domain spectroscopy in the spectral range from 200 GHz to 1 THz was carried out on bulk crystals along different orientations. Principal directions for permittivity were determined along crystallographic axes $\mathbf{c}$, and $\mathbf{b}$, and reciprocal lattice direction $\mathbf{a}^{*}$. No significant frequency dispersion in the real part of dielectric permittivity was observed in the measured spectral range. Our results are in excellent agreement with recent radio-frequency capacitance measurements as well as with extrapolations from recent infrared measurements of phonon mode and high frequency contributions, and close the knowledge gap for these parameters in the terahertz spectral range. Our results are important for applications of $\beta$-Ga$_2$O$_3$ in high-frequency electronic devices. 
\end{abstract}

\maketitle
Single crystalline monoclinic structure beta gallium oxide ($\beta$-Ga$_2$O$_3$), an ultra-wide bandgap semiconductor with direct band gap reported in the range from 4.8~eV to 5.04~eV,\cite{JanowitzNJP2011,Sturm_2016,Sturm_2015,Furthmuller_2016,Mock_2017Ga2O3,GalzkaSST2018} is being extensively researched due to its potential for improved performance in a wide variety of power switching applications as well as radio frequency (RF) components typically employed in power supplies, radar, communication systems, etc.\cite{tsao2018ultrawide} Because of its high estimated breakdown field (E\textsubscript{br}) of approximately 6-8 MV/cm,\cite{higashiwaki2012gallium, yan2018high, pearton2018perspective} its Baliga figure of merit, which is proportional to the third power of $E_{\mathrm{br}}$, has been estimated to be greater than that of Si and GaN.\cite{higashiwaki2012gallium, higashiwaki2016recent} In addition to this primary area of application, since its optical absorption edge lies at ultra-violet (UV) wavelengths, $\beta$-Ga$_2$O$_3$ may also potentially enable the development of solar-blind UV optoelectronic devices\cite{razeghi1996semiconductor, chen2019review} Moreover, a large bandgap energy and controllable $n$-type doping could also facilitate its use as a transparent conductive film for deep-UV applications.\cite{orita2000deep, JanowitzNJP2011} While the overall focus of $\beta$-Ga$_2$O$_3$ research is targeted towards electronic device applications, there is a considerable need to understand its dielectric permittivity $\epsilon(\omega)$ at different frequencies from static (DC), through the far infrared (FIR) and infrared (IR) to the visible,  ultraviolet (UV) and deep ultraviolet (DUV) spectral regions.\cite{fiedler2019static,SchubertPRB2016,bhaumik2011temperature,onuma2016temperature,Sturm_2015,Sturm_2016, Mock_2017Ga2O3}  A fundamental constitutive materials parameter, the permittivity is a direct measure of the polarizability of the constituent atomic lattice and the electron charge distribution under an externally applied electric field. The low-symmetry ($C$2/m) of monoclinic $\beta$-Ga$_2$O$_3$ results in strongly anisotropic dielectric permittivity parameters along different crystal directions. This anisotropic behavior extends over different frequency regions and hence requires comprehensive characterization. Recently, FIR and IR ellipsometric measurements have identified transverse (TO) and longitudinal (LO) phonon mode parameters and their contributions to the anisotropic permittivity in the FIR-IR spectral region. It was observed that phonons polarized within the monoclinic plane do not coincide with specific crystallographic directions, and as a consequence, TO and LO phonons do not coincide in polarization direction either.\cite{SchubertPRB2016,MockPRB2017CWO,MockPRB2018YSO,SchubertPRB2019,StokeyJAP2020LSO} A generalization of the well-known Lyddane-Sachs-Teller relationship was reported for monoclinic and triclinic materials, which relate static and high-frequency permittivity values to the long wavelength active phonon mode parameters.\cite{SchubertPRB2016,SchubertPRL2016GLST} More recently, Fiedler~\textit{et al.} using radio-frequency (RF) measurements on capacitor structures fabricated perpendicular to principal planes (100), (010), and (001) reported strong anisotropy of the RF permittivity.\cite{fiedler2019static} However, in low symmetry crystal systems principal lattice planes are not perpendicular to principal unit cell directions, and the RF measurements did not provide the full set of intrinsic permittivity values yet. In this letter, we characterize the quasi-static dielectric permittivity of $\beta$-Ga$_2$O$_3$ in the frequency range approximately from 0.2-1 THz, and we obtain the full set of all permittivity values for all major lattice directions and directions perpendicular to all principal planes. We compare our results with those from RF and FIR-IR spectral investigations. Our investigation bridges the previously reported FIR-IR and RF studies, and provides information of the dielectric permittivity in the upper gigahertz spectral region of this emerging semiconductor.

Due to the monoclinic symmetry, the anisotropy of $\beta$-Ga$_2$O$_3$ requires careful consideration of lattice directions, and unambiguous assignment of lattice axes and coordinate systems. In general, the dielectric tensor possess four independent complex-valued elements, and cannot be diagonalized for all wavelengths due to frequency dependence of the permittivity. Optical axes, which describe directions within a given crystal along which light can propagate maintaining its polarization, change angular orientation in spectral regions where the permittivity undergoes dispersion in monoclinic crystal systems. This phenomena is also known as rotation of optical axes. This was recently demonstrated for the IR and FIR range for $\beta$-Ga$_2$O$_3$, where all 4 tensor components show strong dispersion behavior across the phonon mode spectral region.\cite{SchubertPRB2016} In another example, Yiwen, Jiyong and Li studied propagation of terahertz waves in a monoclinic crystal of BaGa$_4$Se$_7$.\cite{yiwen2018propagation} The existence of free charge carriers with direction dependent mobility parameters produce anisotropic properties with strong dispersion in the terahertz spectral range. If free charge carriers can be suppressed, either by compensation doping or by using high quality crystalline material, then the dielectric response is composed of all higher frequency bound excitation, such as phonons, band-to-band transitions, excitons, and x-ray absorption. Phonon modes in this material system  are known to be present at frequencies > 3 THz\cite{Ghosh2016_Ab_initio, SchubertPRB2016,Kang2017_Fund_limits}. If phonon modes are far enough away from the terahertz spectral range measured, then dispersion caused by phonons can be neglected. If there is no other dielectric mechanism such as polaron absorption, or charge re-localization between different defects, then the dielectric permittivity tensor should be constant and reveal no significant frequency dispersion. This criterion is important for determining dielectric constants which can then be compared with static or RF measurements. We therefore refer to the terahertz dielectric permittivity parameters here discussed as quasi-static permittivity parameters. We use here polarization-resolved frequency domain spectroscopy methods and extract the anisotropic dielectric permittivity along different crystal directions. We compare our results with recent RF capacitance measurements as well as with extrapolations from recent FIR-IR measurements of phonon mode and high frequency permittivity contributions, and close the knowledge gap for these parameters in the terahertz spectral range. The results of our study are of great technological interest for applications of $\beta$-Ga$_2$O$_3$ in  electronic devices.


\begin{figure}[!htbp]
    \centering
    \includegraphics[width=0.8\columnwidth,keepaspectratio,]{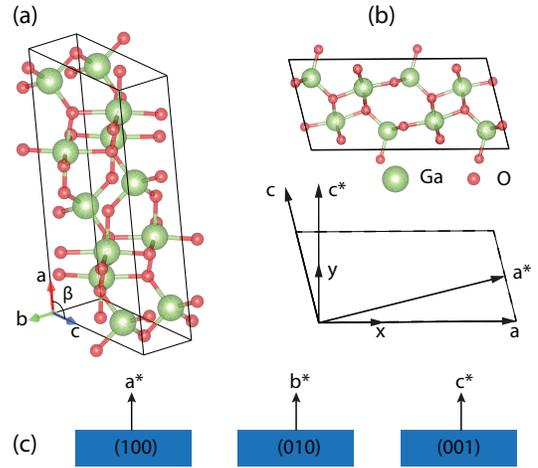}
    \caption{(a) Definition of Cartesian coordinate system ($x$, $y$, $z$), and the unit cell of $\beta$-Ga$_2$O$_3$ with monoclinic angle $\beta$, and crystal unit axes $\mathbf{a}$, $\mathbf{b}$, $\mathbf{c}$. (b) Monoclinic plane $\mathbf{a}$ - $\mathbf{c}$ viewed along axis $\mathbf{b}$. ($\mathbf{b}$ points into the plane.) Reciprocal lattice vectors $\mathbf{c^{\star}}$ parallel to axis $y$ and $\mathbf{a^{\star}}$ parallel to axis $x$ (not drawn to scale). Principal plane Miller indices for surfaces of samples investigated here are shown in (c), with surface normal reciprocal lattice vectors. Note that $\mathbf{b^{\star}}||\mathbf{b}$. The figure was adapted from Ref.\cite{SchubertPRB2016}}
    \label{fig:Ga2O3unitcell}
\end{figure}

Figure~\ref{fig:Ga2O3unitcell} depicts the unit cell of $\beta$-Ga$_2$O$_3$ and a set of Cartesian coordinates, ($x,y,z$). Due to the monoclinic angle ($\beta=103.7^{\circ}$),\cite{GellerJPC1960Ga2O3} axes $\mathbf{a}$ and $\mathbf{c}$ are not perpendicular to each other. The samples investigated in this work are characterized by the Miller indices of the crystallographic planes at the surface. The vectors normal to the planes are obtained by the reciprocal lattice vectors
\begin{equation}
\mathbf{a^{\star}}=\frac{\mathbf{b}\times\mathbf{c}}{\mathbf{a}(\mathbf{b}\times\mathbf{c})},\mbox{ } \mathbf{b^{\star}}=\frac{\mathbf{a}\times\mathbf{c}}{\mathbf{b}(\mathbf{a}\times\mathbf{c})}, \mbox{ } \mathbf{c^{\star}}=\frac{\mathbf{a}\times\mathbf{b}}{\mathbf{c}(\mathbf{a}\times\mathbf{b})},   
\end{equation}
where $\times$ is the vector product, and because $\mathbf{b}$ is perpendicular to $\mathbf{a}$ and $\mathbf{c}$, vector $\mathbf{b^{\star}}$ and $\mathbf{b}$ are parallel. Vectors $\mathbf{a^{\star}}$ and $\mathbf{c^{\star}}$ are depicted in Fig.~~\ref{fig:Ga2O3unitcell}(b).

The frequency dependent permittivity of $\beta$-Ga$_2$O$_3$ can be expressed by a second-rank tensor\cite{SchubertPRB2016}
\begin{equation}\label{eq:epsilontensor}
\varepsilon (\omega)
 =
\begin{bmatrix}
\varepsilon_{xx} & \varepsilon_{xy} & 0\\
\varepsilon_{xy} & \varepsilon_{yy} & 0\\
0&0&\varepsilon_{zz}
   \end{bmatrix},
\end{equation}
where elements $\varepsilon_{xx}$, $\varepsilon_{xy}$, $\varepsilon_{yy}$, and $\varepsilon_{zz}$ are generally independent functions of frequency. With the coordinates defined in Fig.~~\ref{fig:Ga2O3unitcell}, $\varepsilon_{xx}$, $\varepsilon_{yy}$, and $\varepsilon_{zz}$ correspond to permittivities for displacements occurring along axes $\mathbf{a}$, $\mathbf{c^{\star}}$, and $\mathbf{b}$, respectively, while $\varepsilon_{xy}$ corresponds to the shear contribution due to displacements occurring along directions $y$ under electric fields along $x$, and vice versa. Simple Euler angle rotations around axes $z$ ($\phi$), $x$ ($\theta$), and $z'$ ($\psi$; axis $z$ after rotation by $\theta$) as defined in Ref.~\onlinecite{SchubertPRB2016} permit us to express the permittivity tensor in the coordinate system of the measurement setup and according to the crystallographic surface orientation of a given sample. Rotation of $\varepsilon$ in Eq.~\ref{eq:epsilontensor} by $\phi=-(\beta-90^{\circ})$ results in new elements $\varepsilon_{x'x'}$, and $\varepsilon_{y'y'}$ which correspond to permittivity parameters for displacements occurring along axes $\mathbf{a}^{\star}$, and $\mathbf{c}$, respectively,
We determine the anisotropy and the frequency dependent permittivity tensor by frequency-domain polarized terahertz transmission measurements (THz-pT) at normal incidence, and generalized spectroscopic ellipsometry (THz-GSE) at oblique angle of incidence in reflection. 

\begin{figure}
        \includegraphics[width=\columnwidth, keepaspectratio = true]{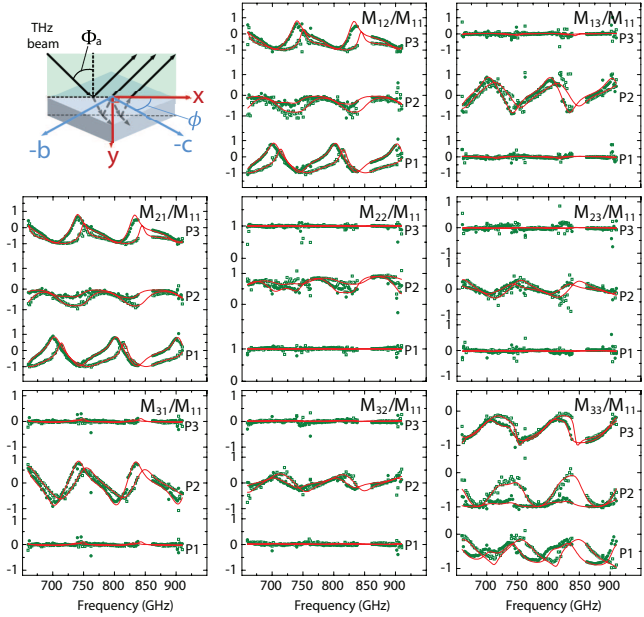}
    \caption{\label{fig:GSE} $\beta$-Ga$_2$O$_3$ (100) GSE data at $\Phi_a=40^{\circ}$ and $60^{\circ}$ angle of incidence:    Dotted green lines (experiment); Solid red lines (best match model calculated). Data are presented in the Mueller matrix formalism. All data are normalized to element M$_{11}$. Data are shown for three azimuths: P1 [$\varphi=-1.5\pm0.2^{\circ}$]; P2 [$\varphi=43.5\pm0.2^{\circ}$]; P3 [$\varphi=88.5\pm0.2^{\circ}$]. The inset depicts schematically the location of axes $\mathbf{b}$ and $\mathbf{c}$ in position P2. Axes $x$ and $y$ indicate the ellipsometer coordinate axes, where $x$ is parallel to the plane of incidence. The sample thickness was determined from the same model analysis as $d=469\pm1\mu m$.}
    \label{THz_MM_(100)}
\end{figure}

We use the Mueller matrix concept\cite{MuellerMIT1943} to describe the electromagnetic properties of an arbitrary anisotropic sample, represented by its input Stokes vector into its outgoing Stokes vector, 
\begin{equation}
\label{eq:muellermatrix}
\left( {{\begin{array}{*{20}c}
 {S_{0} } \hfill \\ {S_{1} } \hfill \\  {S_{2} } \hfill \\  {S_{3} } \hfill \\
\end{array} }} \right)_{\mathrm{output}} =
\left( { { \begin{matrix} { M_{ 11 } } & M_{ 12 } & M_{ 13 } & M_{ 14 } \\ M_{ 21 } & M_{ 22 } & M_{ 23 } & M_{ 24 } \\ M_{ 31 } & M_{ 32 } & M_{ 33 } & M_{ 34 } \\ M_{ 41 } & M_{ 42 } & M_{ 43 } & M_{ 44 } \end{matrix} } } \right) 
\left( {{\begin{array}{*{20}c}
 {S_{0} } \hfill \\ {S_{1} } \hfill \\  {S_{2} } \hfill \\  {S_{3} } \hfill \\
\end{array} }} \right)_{\mathrm{input}}.
\end{equation}
with Stokes vector components defined here by $S_{0}=I_{p}+I_{s}$, $S_{1}=I_{p} - I_{s}$, $S_{2}=I_{45}-I_{ -45}$, $S_{3}=I_{ + }-I_{ - }$, and $I_{p}$, $I_{s}$, $ I_{45}$, $I_{-45}$, $I_{ + }$, and $I_{-}$ denote the intensities for the $p$-, $s$-, +45$^{\circ}$, -45$^{\circ}$, left handed, and right handed circularly polarized light components, respectively.~\cite{SchubertIRSEBook_2004,Fujiwara_2007} In our THz-GSE setup,\cite{KuehneRSI_2014} we obtain elements in the upper 3$\times$3 block normalized to element $M_{11}$. We use matrix algebra approaches to calculate the Mueller matrix elements for arbitrary anisotropic permittivity configurations, as explained in detail previously.\cite{SchubertIRSEBook_2004,SchubertPRB2016,MockPRB2017CWO,Mock_2017Ga2O3,Schubert04} In THz-pT, we measure $M_{11}+M_{12}$, which corresponds to the linearly polarized (horizontal) Stokes vector intensity. 


Three electrically insulating samples were studied in our measurements; (i) Sample A - Czochralski grown Al$^{3+}$ doped crystal with (100) surface orientation \cite{GALAZKA201882,GALAZKA2020125297}, (ii) Sample B- (010) Fe doped crystal and, (iii) Sample C - Fe doped (001) substrate (refer to the Supplementary Information for additional details). The thickness parameters of all samples were determined by mechanical measurements and from analysis of the spectroscopic measurements (in good agreement). Crystallographic plane Miller indices for surfaces of samples investigated here are  shown in Fig.~\ref{fig:Ga2O3unitcell}(c), with surface normal vectors expressed in reciprocal lattice vectors. 

THz-GSE measurements were carried out at $40^{\circ}$ and $60^{\circ}$ angle of incidence in reflection configuration on Sample A and Sample B, wherein the samples were also rotated to multiple azimuth orientations as described in Ref.~\cite{SchubertPRB2016}. THz-GSE data from these samples were analyzed simultaneously in a best-match model regression analysis to determine the values of the relevant permittivity tensor elements. Specific sample orientations were accounted for by proper mathematical rotations of the permittivity tensor. THz-pT measurements were performed on samples A and C using a linearly polarized incident beam at normal incidence. Samples were oriented such that the incident electric field is parallel to $\mathbf{a}^{\star}$, $\mathbf{b}$, and $\mathbf{c}$ axes. 

\begin{figure}
    \centering
    \includegraphics[width=\columnwidth,keepaspectratio = true, scale = 0.5]{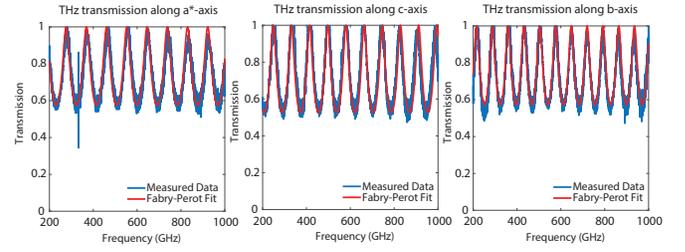}
    \caption{Experimental (symbols) and best-match calculated (lines) THz-pT (polarized transmission) measured for electric field directions very close to (a) $\mathbf{a}^{\star}$, (b) $\mathbf{c}$, and (c) $\mathbf{b}$-axis. The variation in the Fabry-P\'{e}rot fringes between spectra shown in (a) and (b) reveal the strong anisotropy between the permittivities for directions $\mathbf{a}^{\star}$ and $\mathbf{c}$. The sample thickness for the fits in (a) and (b) was found to be 515 $\mu$m while that in (c) is taken to be 656 $\mu$m}
    \label{fig:THzpT}
\end{figure}



Figure~\ref{fig:GSE} depicts selected reflection-type GSE data from sample A, (100) $\beta$-Ga$_2$O$_3$, at multiple angles of incidence, and at multiple sample azimuths. Experimental data and best-match model calculated data are in excellent agreement. Upon rotation to various azimuths, the sample reveals strong anisotropy, recognizable in non-vanishing off-diagonal block elements $M_{31}$, $M_{13}$, $M_{32}$, and $M_{23}$. At azimuth positions P1 and P3, axes $\mathbf{b}$ and $\mathbf{c}$ align with the instrument coordinates $x$ and $y$, and therefore the off-diagonal block elements vanish. Similar data are shown in the Supplementary for sample B. Both data sets were analyzed simultaneously, where parameters for $\varepsilon$ and azimuth orientation $\phi$ are determined. Our analysis reveals that the fourth tensor element $\varepsilon_{xy}$ is too small to be identified in our data analysis. This observation is in agreement with our previous IR-FIR GSE results where the DC value of $\varepsilon_{xy}$ was predicted by extrapolation to be very small (-0.13).\cite{SchubertPRB2016} Hence, In the measured frequency range, we find the permittivity tensor can be diagonalized, where only three elements are needed. These three elements correspond to the principal dielectric axes. An excellent match to the measured Mueller matrix data can be achieved by setting the directions of the principal axes to be along $\mathbf{a}^{\star}$, $\mathbf{b}$, and $\mathbf{c}$. Therefore, the Euler angles for the (100) cut sample A are fixed at $\theta=90^{\circ}$ and $\psi=90^{\circ}$, and for (010) cut sample B to $\theta=0^{\circ}$ and $\psi=0^{\circ}$. We note that the principal dielectric axes directions in monoclinic materials do not necessarily have to coincide with crystallographic axes. For the case of $\beta$-Ga$_2$O$_3$ we find these axes are indistinguishably close to $\mathbf{a}^{\star}$, $\mathbf{b}$, and $\mathbf{c}$. This observation is further verified by the THz-pT measurements as discussed below. Furthermore, we observe no dispersion in the measured frequency range (~0.6-0.9 THz). 

\begin{center}
\begin{table*}[!htbp]
\caption{Static permittivity parameters for $\beta$-Ga$_2$O$_3$ for dielectric displacement along axes $\mathbf{a}$, $\mathbf{a}^{\star}$, $\mathbf{b}$, $\mathbf{c}$, and $\mathbf{c}^{\star}$ as defined in Fig.~\ref{fig:Ga2O3unitcell}, obtained from THz-GSE and THz-pT experiments, in comparison with theory, RF capacitance measurements, and extrapolations from FIR-IR GSE using the eigendielectric polarization model. Note, the values given for the THz-GSE determined $\mathbf{a}$ and $\mathbf{c}^{\star}$ are calculated by rotating the diagonalized permittivity tensor (which contains the values of $\mathbf{a}^{\star}$, $\mathbf{b}$, and $\mathbf{c}$).}
\begin{ruledtabular}
\begin{tabular}{{l}{c}{c}{c}{c}{c}}
    & $\mathbf{a}$&$\mathbf{c}^{\star}$&$\mathbf{a}^{\star}$&$\mathbf{c}$&$\mathbf{b}$ \\
\hline  
THz-GSE & 10.19$^a$ &12.27$^a$&10.05$\pm$0.05&12.4$\pm$0.06&10.6 $\pm$0.06 \\
THz-pT &10.11$^b$&12.21$^b$&9.97$\pm$0.57&12.35$\pm$0.70&10.09 $\pm$0.23\\
RF Capacitance\cite{fiedler2019static} &-&12.4$\pm$0.04&10.2$\pm$0.2&-&10.87$\pm$0.08\\
FIR-IR GSE\cite{SchubertPRB2016} &10.(9)&12.(7)&10.(8)&12.(6)&11.(2)\\
Theory\cite{liu2007lattice} &10.84&-&-&13.89&11.49\\
\end{tabular}
\end{ruledtabular}
\label{tab:epsDC}
$^a$ The resulting off-diagonal tensor component associated with $\mathbf{a}$ and $\mathbf{c}^{\star}$ is -0.54.\\
$^b$ The resulting off-diagonal tensor component associated with $\mathbf{a}$ and $\mathbf{c}^{\star}$ is -0.56.\\
\end{table*}
\end{center}

From the frequency-independent values of $\varepsilon$, permittivity parameters for all crystallographic axes are obtained, and listed in Table.~\ref{tab:epsDC}.
 
 
Figure~\ref{fig:THzpT} depicts THz-pT data for sample B, (010) $\beta$-Ga$_2$O$_3$, and sample C, (001) $\beta$-Ga$_2$O$_3$. The sample orientation was rotated with respect to the polarization of the incident terahertz beam and the corresponding transmission was measured at several angles in the $\mathbf{a}$-$\mathbf{c}$ plane. Given the insulating nature of the Fe-doped substrates, we expect to observe single-mode Fabry-P\'{e}rot (FP) oscillations when the incident E-field is along a major polarizability axis with the transmission levels at the FP maxima approaching near unity. Our measurement of polarization-resolved transmission at different angles in the $\mathbf{a}$-$\mathbf{c}$ plane indicate that the principal axes are close to $\mathbf{a}^{\star}$ and $\mathbf{c}$-directions in the (010) substrate (in agreement with the THz-GSE results). Fig.~\ref{fig:THzpT}(a) and (b) show the transmitted E-field amplitude when the directions $\mathbf{a}^{\star}$ and $\mathbf{c}$ are aligned with the incident polarization, respectively. Fig.~\ref{fig:THzpT}(c) depicts the transmission of the (001) surface with the $\mathbf{b}$ axis parallel to the incident polarization. Experimental data and best-match model are in good agreement (see Supplementary Information for further details). The index of refraction values extracted from these particular analyses are $n_{a*} = 3.16 \pm 0.09 $, $n_{c} = 3.51 \pm 0.10 $ and $n_{b} = 3.18 \pm 0.035 $. No frequency dispersion is observed as evidenced by a constant free spectral range over the measured spectral range. A small drop of the peak transmission levels is observed at the higher frequency end of the measured spectrum (i.e. above 0.7 THz), which, in all cases, is associated with an imaginary part of refractive index $\ll$ 0.01. 


All permittivity values obtained here are listed in Table.~\ref{tab:epsDC} and compared with previous RF and FIR investigation results. Our results are consistent among our different techniques and compare very well with the  previous electrical and optical methods. The RF measurements report permittivity values which correspond to the reciprocal directions of the principal lattice planes, and (100), (010), and (001) planes, respectively, $\mathbf{a^{\star}}$, $\mathbf{b^{\star}}$, and $\mathbf{c^{\star}}$, were investigated. Some of us reported previously on FIR-IR GSE analysis introducing an eigendielectric polarization model for phonon mode contributions to best-match model parameterize the four functions of the permittivity tensor. From the extrapolation in this model towards zero frequency, the DC permittivity values were reported. It is noted here that due to a misprint, values reported in Ref.~\onlinecite{SchubertPRB2016} for permittivity parameters corresponding to $\mathbf{a}$ and $\mathbf{c}^{\star}$ were accidentally switched, and are reported correctly here. We also include values corresponding to $\mathbf{a}^{\star}$ and $\mathbf{c}$, which change very little and within the uncertainty limits. In Table I, it is important to note that with THz-pT measurements we extract the dielectric permittivity values along $\mathbf{a^{\star}}$, $\mathbf{c}$, and $\mathbf{b}$ and values for $\mathbf{a}$ and $\mathbf{c^{\star}}$ are obtained by mathematical rotation of the tensor. The knowledge of the principal directions for permittivity as reported here will be of importance for the accurate simulation of electric field distribution in electronic devices. Furthermore, our results could invariably assist subsequent investigations of terahertz conductivity of epitaxially grown conductive films (e.g., doped $\beta$-Ga$_2$O$_3$). 


In conclusion, we determined the dielectric permittivity along principal lattice directions and directions perpendicular to principal planes for $\beta$-Ga$_2$O$_3$ for 200~GHz to 1~THz. In this spectral range we do not observe any significant dispersion in the real part of the dielectric permittivity. Our results are in good agreement with recent RF and FIR-IR measurements. Our data can be used to predict the permittivity along any direction within the highly anisotropic crystal. The data reported here will be of importance for electronic device designs and simulations involving $\beta$-Ga$_2$O$_3$.

\begin{acknowledgments}
This work was supported in part by the Air Force Office of Scientific Research under awards FA9550-18-1-0507, FA9550-18-1-0360, and FA9550-18-1-0332, by the National Science Foundation under awards DMR 1808715 and ECCS 1810096, by the National Science Foundation supported Nebraska Materials Research Science and Engineering Center under award DMR 1420645 and by the Swedish Governmental Agency for Innovation Systems (VINNOVA) under the Competence Center Program Grant No. 2016–05190, the Swedish Research Council VR award No. 2016-00889, the Swedish Foundation for Strategic Research Grant Nos.  RIF14-055, EM16-0024, by the Knut and Alice Wallenbergs Foundation supported grant 'Wide-bandgap semiconductors for next generation quantum components', and by the Swedish Government Strategic Research Area in Materials Science on Functional Materials at Link{\"o}ping University, Faculty Grant SFO Mat LiU No. 2009-00971. This work was also partly performed in the frame-work of GraFOx, a Leibniz-Science Campus partially funded by the Leibniz Association – Germany. M. S. acknowledges the University of Nebraska Foundation and the J. A. Woollam Foundation for financial support. The authors thank John Blevins (Air Force Research Lab) for providing the Synoptics (010)  $\beta$-Ga$_2$O$_3$ bulk substrates used in the measurements.\\
\end{acknowledgments}

The data that support the findings of this study are available from the corresponding author upon reasonable request.

\nocite{*}
\bibliography{aipsamp}

\begin{thebibliography}{39}%
\makeatletter
\providecommand \@ifxundefined [1]{%
 \@ifx{#1\undefined}
}%
\providecommand \@ifnum [1]{%
 \ifnum #1\expandafter \@firstoftwo
 \else \expandafter \@secondoftwo
 \fi
}%
\providecommand \@ifx [1]{%
 \ifx #1\expandafter \@firstoftwo
 \else \expandafter \@secondoftwo
 \fi
}%
\providecommand \natexlab [1]{#1}%
\providecommand \enquote  [1]{``#1''}%
\providecommand \bibnamefont  [1]{#1}%
\providecommand \bibfnamefont [1]{#1}%
\providecommand \citenamefont [1]{#1}%
\providecommand \href@noop [0]{\@secondoftwo}%
\providecommand \href [0]{\begingroup \@sanitize@url \@href}%
\providecommand \@href[1]{\@@startlink{#1}\@@href}%
\providecommand \@@href[1]{\endgroup#1\@@endlink}%
\providecommand \@sanitize@url [0]{\catcode `\\12\catcode `\$12\catcode
  `\&12\catcode `\#12\catcode `\^12\catcode `\_12\catcode `\%12\relax}%
\providecommand \@@startlink[1]{}%
\providecommand \@@endlink[0]{}%
\providecommand \url  [0]{\begingroup\@sanitize@url \@url }%
\providecommand \@url [1]{\endgroup\@href {#1}{\urlprefix }}%
\providecommand \urlprefix  [0]{URL }%
\providecommand \Eprint [0]{\href }%
\providecommand \doibase [0]{http://dx.doi.org/}%
\providecommand \selectlanguage [0]{\@gobble}%
\providecommand \bibinfo  [0]{\@secondoftwo}%
\providecommand \bibfield  [0]{\@secondoftwo}%
\providecommand \translation [1]{[#1]}%
\providecommand \BibitemOpen [0]{}%
\providecommand \bibitemStop [0]{}%
\providecommand \bibitemNoStop [0]{.\EOS\space}%
\providecommand \EOS [0]{\spacefactor3000\relax}%
\providecommand \BibitemShut  [1]{\csname bibitem#1\endcsname}%
\let\auto@bib@innerbib\@empty
\bibitem [{\citenamefont {Janowitz}\ \emph {et~al.}(2011)\citenamefont
  {Janowitz}, \citenamefont {Scherer}, \citenamefont {Mohamed}, \citenamefont
  {Krapf}, \citenamefont {Dwelk}, \citenamefont {Manzke}, \citenamefont
  {Galazka}, \citenamefont {Uecker}, \citenamefont {Irmscher}, \citenamefont
  {Fornari}, \citenamefont {Michling}, \citenamefont {Schmei{\ss}er},
  \citenamefont {Weber}, \citenamefont {Varley},\ and\ \citenamefont
  {de~Walle}}]{JanowitzNJP2011}%
  \BibitemOpen
  \bibfield  {author} {\bibinfo {author} {\bibfnamefont {C.}~\bibnamefont
  {Janowitz}}, \bibinfo {author} {\bibfnamefont {V.}~\bibnamefont {Scherer}},
  \bibinfo {author} {\bibfnamefont {M.}~\bibnamefont {Mohamed}}, \bibinfo
  {author} {\bibfnamefont {A.}~\bibnamefont {Krapf}}, \bibinfo {author}
  {\bibfnamefont {H.}~\bibnamefont {Dwelk}}, \bibinfo {author} {\bibfnamefont
  {R.}~\bibnamefont {Manzke}}, \bibinfo {author} {\bibfnamefont
  {Z.}~\bibnamefont {Galazka}}, \bibinfo {author} {\bibfnamefont
  {R.}~\bibnamefont {Uecker}}, \bibinfo {author} {\bibfnamefont
  {K.}~\bibnamefont {Irmscher}}, \bibinfo {author} {\bibfnamefont
  {R.}~\bibnamefont {Fornari}}, \bibinfo {author} {\bibfnamefont
  {M.}~\bibnamefont {Michling}}, \bibinfo {author} {\bibfnamefont
  {D.}~\bibnamefont {Schmei{\ss}er}}, \bibinfo {author} {\bibfnamefont {J.~R.}\
  \bibnamefont {Weber}}, \bibinfo {author} {\bibfnamefont {J.~B.}\ \bibnamefont
  {Varley}}, \ and\ \bibinfo {author} {\bibfnamefont {C.~G.~V.}\ \bibnamefont
  {de~Walle}},\ }\href@noop {} {\bibfield  {journal} {\bibinfo  {journal} {New
  J. Phys.}\ }\textbf {\bibinfo {volume} {13}},\ \bibinfo {pages} {085014}
  (\bibinfo {year} {2011})}\BibitemShut {NoStop}%
\bibitem [{\citenamefont {Sturm}\ \emph {et~al.}(2016)\citenamefont {Sturm},
  \citenamefont {Schmidt-Grund}, \citenamefont {Kranert}, \citenamefont
  {Furthm{\"u}ller}, \citenamefont {Bechstedt},\ and\ \citenamefont
  {Grundmann}}]{Sturm_2016}%
  \BibitemOpen
  \bibfield  {author} {\bibinfo {author} {\bibfnamefont {C.}~\bibnamefont
  {Sturm}}, \bibinfo {author} {\bibfnamefont {R.}~\bibnamefont
  {Schmidt-Grund}}, \bibinfo {author} {\bibfnamefont {C.}~\bibnamefont
  {Kranert}}, \bibinfo {author} {\bibfnamefont {J.}~\bibnamefont
  {Furthm{\"u}ller}}, \bibinfo {author} {\bibfnamefont {F.}~\bibnamefont
  {Bechstedt}}, \ and\ \bibinfo {author} {\bibfnamefont {M.}~\bibnamefont
  {Grundmann}},\ }\bibfield  {title} {\enquote {\bibinfo {title} {Dipole
  analysis of the dielectric function of colour dispersive materials:
  Application to monoclinic \uppercase{G}a$_2$\uppercase{O}$_3$},}\ }\href@noop
  {} {\bibfield  {journal} {\bibinfo  {journal} {Phys. Rev. B}\ }\textbf
  {\bibinfo {volume} {94}},\ \bibinfo {pages} {035148} (\bibinfo {year}
  {2016})}\BibitemShut {NoStop}%
\bibitem [{\citenamefont {Sturm}\ \emph {et~al.}(2015)\citenamefont {Sturm},
  \citenamefont {Furthm{\"u}ller}, \citenamefont {Bechstedt}, \citenamefont
  {Schmidt-Grund},\ and\ \citenamefont {Grundmann}}]{Sturm_2015}%
  \BibitemOpen
  \bibfield  {author} {\bibinfo {author} {\bibfnamefont {C.}~\bibnamefont
  {Sturm}}, \bibinfo {author} {\bibfnamefont {J.}~\bibnamefont
  {Furthm{\"u}ller}}, \bibinfo {author} {\bibfnamefont {F.}~\bibnamefont
  {Bechstedt}}, \bibinfo {author} {\bibfnamefont {R.}~\bibnamefont
  {Schmidt-Grund}}, \ and\ \bibinfo {author} {\bibfnamefont {M.}~\bibnamefont
  {Grundmann}},\ }\bibfield  {title} {\enquote {\bibinfo {title} {Dielectric
  tensor of monoclinic \uppercase{G}a$_2$\uppercase{O}$_3$ single crystals in
  the spectral range 0.5--8.5 ev},}\ }\href@noop {} {\bibfield  {journal}
  {\bibinfo  {journal} {APL Mater.}\ }\textbf {\bibinfo {volume} {3}},\
  \bibinfo {pages} {106106} (\bibinfo {year} {2015})}\BibitemShut {NoStop}%
\bibitem [{\citenamefont {Furthm\"uller}\ and\ \citenamefont
  {Bechstedt}(2016)}]{Furthmuller_2016}%
  \BibitemOpen
  \bibfield  {author} {\bibinfo {author} {\bibfnamefont {J.}~\bibnamefont
  {Furthm\"uller}}\ and\ \bibinfo {author} {\bibfnamefont {F.}~\bibnamefont
  {Bechstedt}},\ }\bibfield  {title} {\enquote {\bibinfo {title} {Quasiparticle
  bands and spectra of \uppercase{G}a$_2$\uppercase{O}$_3$ polymorphs},}\
  }\href {\doibase 10.1103/PhysRevB.93.115204} {\bibfield  {journal} {\bibinfo
  {journal} {Phys. Rev. B}\ }\textbf {\bibinfo {volume} {93}},\ \bibinfo
  {pages} {115204} (\bibinfo {year} {2016})}\BibitemShut {NoStop}%
\bibitem [{\citenamefont {Mock}\ \emph
  {et~al.}(2017{\natexlab{a}})\citenamefont {Mock}, \citenamefont {Korlacki},
  \citenamefont {Briley}, \citenamefont {Darakchieva}, \citenamefont {Monemar},
  \citenamefont {Kumagai}, \citenamefont {Goto}, \citenamefont {Higashiwaki},\
  and\ \citenamefont {Schubert}}]{Mock_2017Ga2O3}%
  \BibitemOpen
  \bibfield  {author} {\bibinfo {author} {\bibfnamefont {A.}~\bibnamefont
  {Mock}}, \bibinfo {author} {\bibfnamefont {R.}~\bibnamefont {Korlacki}},
  \bibinfo {author} {\bibfnamefont {C.}~\bibnamefont {Briley}}, \bibinfo
  {author} {\bibfnamefont {V.}~\bibnamefont {Darakchieva}}, \bibinfo {author}
  {\bibfnamefont {B.}~\bibnamefont {Monemar}}, \bibinfo {author} {\bibfnamefont
  {Y.}~\bibnamefont {Kumagai}}, \bibinfo {author} {\bibfnamefont
  {K.}~\bibnamefont {Goto}}, \bibinfo {author} {\bibfnamefont {M.}~\bibnamefont
  {Higashiwaki}}, \ and\ \bibinfo {author} {\bibfnamefont {M.}~\bibnamefont
  {Schubert}},\ }\bibfield  {title} {\enquote {\bibinfo {title} {Band-to-band
  transitions, selection rules, effective mass, and excitonic contributions in
  monoclinic$\beta$-\uppercase{G}a$_2$\uppercase{O}$_3$},}\ }\href@noop {}
  {\bibfield  {journal} {\bibinfo  {journal} {Phys.\ Rev.\ B}\ }\textbf
  {\bibinfo {volume} {96}},\ \bibinfo {pages} {245205} (\bibinfo {year}
  {2017}{\natexlab{a}})}\BibitemShut {NoStop}%
\bibitem [{\citenamefont {Galazka}(2018)}]{GalzkaSST2018}%
  \BibitemOpen
  \bibfield  {author} {\bibinfo {author} {\bibfnamefont {Z.}~\bibnamefont
  {Galazka}},\ }\bibfield  {title} {\enquote {\bibinfo {title}
  {$\beta$-\uppercase{G}a$_2$\uppercase{O}$_3$ for wide-bandgap electronics and
  optoelectronics},}\ }\href@noop {} {\bibfield  {journal} {\bibinfo  {journal}
  {Semicond. Sci. Technol.}\ }\textbf {\bibinfo {volume} {33}},\ \bibinfo
  {pages} {113001} (\bibinfo {year} {2018})}\BibitemShut {NoStop}%
\bibitem [{\citenamefont {Tsao}\ \emph {et~al.}(2018)\citenamefont {Tsao},
  \citenamefont {Chowdhury}, \citenamefont {Hollis}, \citenamefont {Jena},
  \citenamefont {Johnson}, \citenamefont {Jones}, \citenamefont {Kaplar},
  \citenamefont {Rajan}, \citenamefont {Van~de Walle}, \citenamefont {Bellotti}
  \emph {et~al.}}]{tsao2018ultrawide}%
  \BibitemOpen
  \bibfield  {author} {\bibinfo {author} {\bibfnamefont {J.}~\bibnamefont
  {Tsao}}, \bibinfo {author} {\bibfnamefont {S.}~\bibnamefont {Chowdhury}},
  \bibinfo {author} {\bibfnamefont {M.}~\bibnamefont {Hollis}}, \bibinfo
  {author} {\bibfnamefont {D.}~\bibnamefont {Jena}}, \bibinfo {author}
  {\bibfnamefont {N.}~\bibnamefont {Johnson}}, \bibinfo {author} {\bibfnamefont
  {K.}~\bibnamefont {Jones}}, \bibinfo {author} {\bibfnamefont
  {R.}~\bibnamefont {Kaplar}}, \bibinfo {author} {\bibfnamefont
  {S.}~\bibnamefont {Rajan}}, \bibinfo {author} {\bibfnamefont
  {C.}~\bibnamefont {Van~de Walle}}, \bibinfo {author} {\bibfnamefont
  {E.}~\bibnamefont {Bellotti}},  \emph {et~al.},\ }\bibfield  {title}
  {\enquote {\bibinfo {title} {Ultrawide-bandgap semiconductors: Research
  opportunities and challenges},}\ }\href@noop {} {\bibfield  {journal}
  {\bibinfo  {journal} {Advanced Electronic Materials.}\ }\textbf {\bibinfo
  {volume} {4}},\ \bibinfo {pages} {1600501} (\bibinfo {year}
  {2018})}\BibitemShut {NoStop}%
\bibitem [{\citenamefont {Higashiwaki}\ \emph {et~al.}(2012)\citenamefont
  {Higashiwaki}, \citenamefont {Sasaki}, \citenamefont {Kuramata},
  \citenamefont {Masui},\ and\ \citenamefont
  {Yamakoshi}}]{higashiwaki2012gallium}%
  \BibitemOpen
  \bibfield  {author} {\bibinfo {author} {\bibfnamefont {M.}~\bibnamefont
  {Higashiwaki}}, \bibinfo {author} {\bibfnamefont {K.}~\bibnamefont {Sasaki}},
  \bibinfo {author} {\bibfnamefont {A.}~\bibnamefont {Kuramata}}, \bibinfo
  {author} {\bibfnamefont {T.}~\bibnamefont {Masui}}, \ and\ \bibinfo {author}
  {\bibfnamefont {S.}~\bibnamefont {Yamakoshi}},\ }\bibfield  {title} {\enquote
  {\bibinfo {title} {Gallium oxide (\uppercase{G}a$_2$\uppercase{O}$_3$)
  metal-semiconductor field-effect transistors on single-crystal
  $\beta$-\uppercase{G}a$_2$\uppercase{O}$_3$ (010) substrates},}\ }\href@noop
  {} {\bibfield  {journal} {\bibinfo  {journal} {Applied Physics Letters}\
  }\textbf {\bibinfo {volume} {100}},\ \bibinfo {pages} {013504} (\bibinfo
  {year} {2012})}\BibitemShut {NoStop}%
\bibitem [{\citenamefont {Yan}\ \emph {et~al.}(2018)\citenamefont {Yan},
  \citenamefont {Esqueda}, \citenamefont {Ma}, \citenamefont {Tice},\ and\
  \citenamefont {Wang}}]{yan2018high}%
  \BibitemOpen
  \bibfield  {author} {\bibinfo {author} {\bibfnamefont {X.}~\bibnamefont
  {Yan}}, \bibinfo {author} {\bibfnamefont {I.~S.}\ \bibnamefont {Esqueda}},
  \bibinfo {author} {\bibfnamefont {J.}~\bibnamefont {Ma}}, \bibinfo {author}
  {\bibfnamefont {J.}~\bibnamefont {Tice}}, \ and\ \bibinfo {author}
  {\bibfnamefont {H.}~\bibnamefont {Wang}},\ }\bibfield  {title} {\enquote
  {\bibinfo {title} {High breakdown electric field in
  $\beta$-\uppercase{G}a$_2$\uppercase{O}$_3$/graphene vertical barristor
  heterostructure},}\ }\href@noop {} {\bibfield  {journal} {\bibinfo  {journal}
  {Applied Physics Letters}\ }\textbf {\bibinfo {volume} {112}},\ \bibinfo
  {pages} {032101} (\bibinfo {year} {2018})}\BibitemShut {NoStop}%
\bibitem [{\citenamefont {Pearton}\ \emph {et~al.}(2018)\citenamefont
  {Pearton}, \citenamefont {Ren}, \citenamefont {Tadjer},\ and\ \citenamefont
  {Kim}}]{pearton2018perspective}%
  \BibitemOpen
  \bibfield  {author} {\bibinfo {author} {\bibfnamefont {S.}~\bibnamefont
  {Pearton}}, \bibinfo {author} {\bibfnamefont {F.}~\bibnamefont {Ren}},
  \bibinfo {author} {\bibfnamefont {M.}~\bibnamefont {Tadjer}}, \ and\ \bibinfo
  {author} {\bibfnamefont {J.}~\bibnamefont {Kim}},\ }\bibfield  {title}
  {\enquote {\bibinfo {title} {Perspective: \uppercase{G}a$_2$\uppercase{O}$_3$
  for ultra-high power rectifiers and mosfets},}\ }\href@noop {} {\bibfield
  {journal} {\bibinfo  {journal} {Journal of Applied Physics}\ }\textbf
  {\bibinfo {volume} {124}},\ \bibinfo {pages} {220901} (\bibinfo {year}
  {2018})}\BibitemShut {NoStop}%
\bibitem [{\citenamefont {Higashiwaki}\ \emph {et~al.}(2016)\citenamefont
  {Higashiwaki}, \citenamefont {Sasaki}, \citenamefont {Murakami},
  \citenamefont {Kumagai}, \citenamefont {Koukitu}, \citenamefont {Kuramata},
  \citenamefont {Masui},\ and\ \citenamefont
  {Yamakoshi}}]{higashiwaki2016recent}%
  \BibitemOpen
  \bibfield  {author} {\bibinfo {author} {\bibfnamefont {M.}~\bibnamefont
  {Higashiwaki}}, \bibinfo {author} {\bibfnamefont {K.}~\bibnamefont {Sasaki}},
  \bibinfo {author} {\bibfnamefont {H.}~\bibnamefont {Murakami}}, \bibinfo
  {author} {\bibfnamefont {Y.}~\bibnamefont {Kumagai}}, \bibinfo {author}
  {\bibfnamefont {A.}~\bibnamefont {Koukitu}}, \bibinfo {author} {\bibfnamefont
  {A.}~\bibnamefont {Kuramata}}, \bibinfo {author} {\bibfnamefont
  {T.}~\bibnamefont {Masui}}, \ and\ \bibinfo {author} {\bibfnamefont
  {S.}~\bibnamefont {Yamakoshi}},\ }\bibfield  {title} {\enquote {\bibinfo
  {title} {Recent progress in \uppercase{G}a$_2$\uppercase{O}$_3$ power
  devices},}\ }\href@noop {} {\bibfield  {journal} {\bibinfo  {journal}
  {Semiconductor Science and Technology}\ }\textbf {\bibinfo {volume} {31}},\
  \bibinfo {pages} {034001} (\bibinfo {year} {2016})}\BibitemShut {NoStop}%
\bibitem [{\citenamefont {Razeghi}\ and\ \citenamefont
  {Rogalski}(1996)}]{razeghi1996semiconductor}%
  \BibitemOpen
  \bibfield  {author} {\bibinfo {author} {\bibfnamefont {M.}~\bibnamefont
  {Razeghi}}\ and\ \bibinfo {author} {\bibfnamefont {A.}~\bibnamefont
  {Rogalski}},\ }\bibfield  {title} {\enquote {\bibinfo {title} {Semiconductor
  ultraviolet detectors},}\ }\href@noop {} {\bibfield  {journal} {\bibinfo
  {journal} {Journal of Applied Physics}\ }\textbf {\bibinfo {volume} {79}},\
  \bibinfo {pages} {7433--7473} (\bibinfo {year} {1996})}\BibitemShut {NoStop}%
\bibitem [{\citenamefont {Chen}\ \emph {et~al.}(2019)\citenamefont {Chen},
  \citenamefont {Ren}, \citenamefont {Gu},\ and\ \citenamefont
  {Ye}}]{chen2019review}%
  \BibitemOpen
  \bibfield  {author} {\bibinfo {author} {\bibfnamefont {X.}~\bibnamefont
  {Chen}}, \bibinfo {author} {\bibfnamefont {F.}~\bibnamefont {Ren}}, \bibinfo
  {author} {\bibfnamefont {S.}~\bibnamefont {Gu}}, \ and\ \bibinfo {author}
  {\bibfnamefont {J.}~\bibnamefont {Ye}},\ }\bibfield  {title} {\enquote
  {\bibinfo {title} {Review of gallium-oxide-based solar-blind ultraviolet
  photodetectors},}\ }\href@noop {} {\bibfield  {journal} {\bibinfo  {journal}
  {Photonics Research}\ }\textbf {\bibinfo {volume} {7}},\ \bibinfo {pages}
  {381--415} (\bibinfo {year} {2019})}\BibitemShut {NoStop}%
\bibitem [{\citenamefont {Orita}\ \emph {et~al.}(2000)\citenamefont {Orita},
  \citenamefont {Ohta}, \citenamefont {Hirano},\ and\ \citenamefont
  {Hosono}}]{orita2000deep}%
  \BibitemOpen
  \bibfield  {author} {\bibinfo {author} {\bibfnamefont {M.}~\bibnamefont
  {Orita}}, \bibinfo {author} {\bibfnamefont {H.}~\bibnamefont {Ohta}},
  \bibinfo {author} {\bibfnamefont {M.}~\bibnamefont {Hirano}}, \ and\ \bibinfo
  {author} {\bibfnamefont {H.}~\bibnamefont {Hosono}},\ }\bibfield  {title}
  {\enquote {\bibinfo {title} {Deep-ultraviolet transparent conductive
  $\beta$-\uppercase{G}a$_2$\uppercase{O}$_3$ thin films},}\ }\href@noop {}
  {\bibfield  {journal} {\bibinfo  {journal} {Applied Physics Letters}\
  }\textbf {\bibinfo {volume} {77}},\ \bibinfo {pages} {4166--4168} (\bibinfo
  {year} {2000})}\BibitemShut {NoStop}%
\bibitem [{\citenamefont {Fiedler}\ \emph {et~al.}(2019)\citenamefont
  {Fiedler}, \citenamefont {Schewski}, \citenamefont {Galazka},\ and\
  \citenamefont {Irmscher}}]{fiedler2019static}%
  \BibitemOpen
  \bibfield  {author} {\bibinfo {author} {\bibfnamefont {A.}~\bibnamefont
  {Fiedler}}, \bibinfo {author} {\bibfnamefont {R.}~\bibnamefont {Schewski}},
  \bibinfo {author} {\bibfnamefont {Z.}~\bibnamefont {Galazka}}, \ and\
  \bibinfo {author} {\bibfnamefont {K.}~\bibnamefont {Irmscher}},\ }\bibfield
  {title} {\enquote {\bibinfo {title} {Static dielectric constant of
  $\beta$-\uppercase{G}a$_2$\uppercase{O}$_3$ perpendicular to the principal
  planes (100),(010), and (001)},}\ }\href@noop {} {\bibfield  {journal}
  {\bibinfo  {journal} {ECS Journal of Solid State Science and Technology}\
  }\textbf {\bibinfo {volume} {8}},\ \bibinfo {pages} {Q3083--Q3085} (\bibinfo
  {year} {2019})}\BibitemShut {NoStop}%
\bibitem [{\citenamefont {Schubert}\ \emph {et~al.}(2016)\citenamefont
  {Schubert}, \citenamefont {Korlacki}, \citenamefont {Knight}, \citenamefont
  {Hofmann}, \citenamefont {Sch\"{o}che}, \citenamefont {Darakchieva},
  \citenamefont {Janz\'{e}n}, \citenamefont {Monemar}, \citenamefont {Gogova},
  \citenamefont {Thieu}, \citenamefont {Togashi}, \citenamefont {Murakami},
  \citenamefont {Kumagai}, \citenamefont {Goto}, \citenamefont {Kuramata},
  \citenamefont {Yamakoshi},\ and\ \citenamefont
  {Higashiwaki}}]{SchubertPRB2016}%
  \BibitemOpen
  \bibfield  {author} {\bibinfo {author} {\bibfnamefont {M.}~\bibnamefont
  {Schubert}}, \bibinfo {author} {\bibfnamefont {R.}~\bibnamefont {Korlacki}},
  \bibinfo {author} {\bibfnamefont {S.}~\bibnamefont {Knight}}, \bibinfo
  {author} {\bibfnamefont {T.}~\bibnamefont {Hofmann}}, \bibinfo {author}
  {\bibfnamefont {S.}~\bibnamefont {Sch\"{o}che}}, \bibinfo {author}
  {\bibfnamefont {V.}~\bibnamefont {Darakchieva}}, \bibinfo {author}
  {\bibfnamefont {E.}~\bibnamefont {Janz\'{e}n}}, \bibinfo {author}
  {\bibfnamefont {B.}~\bibnamefont {Monemar}}, \bibinfo {author} {\bibfnamefont
  {D.}~\bibnamefont {Gogova}}, \bibinfo {author} {\bibfnamefont {Q.-T.}\
  \bibnamefont {Thieu}}, \bibinfo {author} {\bibfnamefont {R.}~\bibnamefont
  {Togashi}}, \bibinfo {author} {\bibfnamefont {H.}~\bibnamefont {Murakami}},
  \bibinfo {author} {\bibfnamefont {Y.}~\bibnamefont {Kumagai}}, \bibinfo
  {author} {\bibfnamefont {K.}~\bibnamefont {Goto}}, \bibinfo {author}
  {\bibfnamefont {A.}~\bibnamefont {Kuramata}}, \bibinfo {author}
  {\bibfnamefont {S.}~\bibnamefont {Yamakoshi}}, \ and\ \bibinfo {author}
  {\bibfnamefont {M.}~\bibnamefont {Higashiwaki}},\ }\bibfield  {title}
  {\enquote {\bibinfo {title} {Anisotropy, phonon modes, and free charge
  carrier parameters in monoclinic $\beta$-gallium oxide single crystals},}\
  }\href@noop {} {\bibfield  {journal} {\bibinfo  {journal} {Phys. Rev. B}\
  }\textbf {\bibinfo {volume} {93}},\ \bibinfo {pages} {125209} (\bibinfo
  {year} {2016})}\BibitemShut {NoStop}%
\bibitem [{\citenamefont {Bhaumik}\ \emph {et~al.}(2011)\citenamefont
  {Bhaumik}, \citenamefont {Bhatt}, \citenamefont {Ganesamoorthy},
  \citenamefont {Saxena}, \citenamefont {Karnal}, \citenamefont {Gupta},
  \citenamefont {Sinha},\ and\ \citenamefont {Deb}}]{bhaumik2011temperature}%
  \BibitemOpen
  \bibfield  {author} {\bibinfo {author} {\bibfnamefont {I.}~\bibnamefont
  {Bhaumik}}, \bibinfo {author} {\bibfnamefont {R.}~\bibnamefont {Bhatt}},
  \bibinfo {author} {\bibfnamefont {S.}~\bibnamefont {Ganesamoorthy}}, \bibinfo
  {author} {\bibfnamefont {A.}~\bibnamefont {Saxena}}, \bibinfo {author}
  {\bibfnamefont {A.}~\bibnamefont {Karnal}}, \bibinfo {author} {\bibfnamefont
  {P.}~\bibnamefont {Gupta}}, \bibinfo {author} {\bibfnamefont
  {A.}~\bibnamefont {Sinha}}, \ and\ \bibinfo {author} {\bibfnamefont
  {S.}~\bibnamefont {Deb}},\ }\bibfield  {title} {\enquote {\bibinfo {title}
  {Temperature-dependent index of refraction of monoclinic
  \uppercase{G}a$_2$\uppercase{O}$_3$ single crystal},}\ }\href@noop {}
  {\bibfield  {journal} {\bibinfo  {journal} {Applied optics}\ }\textbf
  {\bibinfo {volume} {50}},\ \bibinfo {pages} {6006--6010} (\bibinfo {year}
  {2011})}\BibitemShut {NoStop}%
\bibitem [{\citenamefont {Onuma}\ \emph {et~al.}(2016)\citenamefont {Onuma},
  \citenamefont {Saito}, \citenamefont {Sasaki}, \citenamefont {Goto},
  \citenamefont {Masui}, \citenamefont {Yamaguchi}, \citenamefont {Honda},
  \citenamefont {Kuramata},\ and\ \citenamefont
  {Higashiwaki}}]{onuma2016temperature}%
  \BibitemOpen
  \bibfield  {author} {\bibinfo {author} {\bibfnamefont {T.}~\bibnamefont
  {Onuma}}, \bibinfo {author} {\bibfnamefont {S.}~\bibnamefont {Saito}},
  \bibinfo {author} {\bibfnamefont {K.}~\bibnamefont {Sasaki}}, \bibinfo
  {author} {\bibfnamefont {K.}~\bibnamefont {Goto}}, \bibinfo {author}
  {\bibfnamefont {T.}~\bibnamefont {Masui}}, \bibinfo {author} {\bibfnamefont
  {T.}~\bibnamefont {Yamaguchi}}, \bibinfo {author} {\bibfnamefont
  {T.}~\bibnamefont {Honda}}, \bibinfo {author} {\bibfnamefont
  {A.}~\bibnamefont {Kuramata}}, \ and\ \bibinfo {author} {\bibfnamefont
  {M.}~\bibnamefont {Higashiwaki}},\ }\bibfield  {title} {\enquote {\bibinfo
  {title} {Temperature-dependent exciton resonance energies and their
  correlation with ir-active optical phonon modes in
  $\beta$-\uppercase{G}a$_2$\uppercase{O}$_3$ single crystals},}\ }\href@noop
  {} {\bibfield  {journal} {\bibinfo  {journal} {Applied Physics Letters}\
  }\textbf {\bibinfo {volume} {108}},\ \bibinfo {pages} {101904} (\bibinfo
  {year} {2016})}\BibitemShut {NoStop}%
\bibitem [{\citenamefont {Mock}\ \emph
  {et~al.}(2017{\natexlab{b}})\citenamefont {Mock}, \citenamefont {Korlacki},
  \citenamefont {Knight},\ and\ \citenamefont {Schubert}}]{MockPRB2017CWO}%
  \BibitemOpen
  \bibfield  {author} {\bibinfo {author} {\bibfnamefont {A.}~\bibnamefont
  {Mock}}, \bibinfo {author} {\bibfnamefont {R.}~\bibnamefont {Korlacki}},
  \bibinfo {author} {\bibfnamefont {S.}~\bibnamefont {Knight}}, \ and\ \bibinfo
  {author} {\bibfnamefont {M.}~\bibnamefont {Schubert}},\ }\bibfield  {title}
  {\enquote {\bibinfo {title} {Anisotropy, phonon modes, and lattice
  anharmonicity from dielectric function tensor analysis of monoclinic cadmium
  tungstate},}\ }\href@noop {} {\bibfield  {journal} {\bibinfo  {journal}
  {Phys. Rev. B}\ }\textbf {\bibinfo {volume} {95}},\ \bibinfo {pages} {165202}
  (\bibinfo {year} {2017}{\natexlab{b}})}\BibitemShut {NoStop}%
\bibitem [{\citenamefont {Mock}\ \emph {et~al.}(2018)\citenamefont {Mock},
  \citenamefont {Korlacki}, \citenamefont {Knight},\ and\ \citenamefont
  {Schubert}}]{MockPRB2018YSO}%
  \BibitemOpen
  \bibfield  {author} {\bibinfo {author} {\bibfnamefont {A.}~\bibnamefont
  {Mock}}, \bibinfo {author} {\bibfnamefont {R.}~\bibnamefont {Korlacki}},
  \bibinfo {author} {\bibfnamefont {S.}~\bibnamefont {Knight}}, \ and\ \bibinfo
  {author} {\bibfnamefont {M.}~\bibnamefont {Schubert}},\ }\bibfield  {title}
  {\enquote {\bibinfo {title} {Anisotropy and phonon modes from analysis of the
  dielectric function tensor and inverse dielectric function tensor of
  monoclinic yttrium orthosilicate},}\ }\href@noop {} {\bibfield  {journal}
  {\bibinfo  {journal} {Phys. Rev. B}\ }\textbf {\bibinfo {volume} {97}},\
  \bibinfo {pages} {165203} (\bibinfo {year} {2018})}\BibitemShut {NoStop}%
\bibitem [{\citenamefont {Schubert}\ \emph
  {et~al.}(2019{\natexlab{a}})\citenamefont {Schubert}, \citenamefont {Mock},
  \citenamefont {Korlacki},\ and\ \citenamefont
  {Darakchieva}}]{SchubertPRB2019}%
  \BibitemOpen
  \bibfield  {author} {\bibinfo {author} {\bibfnamefont {M.}~\bibnamefont
  {Schubert}}, \bibinfo {author} {\bibfnamefont {A.}~\bibnamefont {Mock}},
  \bibinfo {author} {\bibfnamefont {R.}~\bibnamefont {Korlacki}}, \ and\
  \bibinfo {author} {\bibfnamefont {V.}~\bibnamefont {Darakchieva}},\
  }\bibfield  {title} {\enquote {\bibinfo {title} {Phonon order and
  reststrahlen bands of polar vibrations in crystals with monoclinic
  symmetry},}\ }\href@noop {} {\bibfield  {journal} {\bibinfo  {journal} {Phys.
  Rev. B}\ }\textbf {\bibinfo {volume} {99}},\ \bibinfo {pages} {041201(R)}
  (\bibinfo {year} {2019}{\natexlab{a}})}\BibitemShut {NoStop}%
\bibitem [{\citenamefont {Stokey}\ \emph {et~al.}(2020)\citenamefont {Stokey},
  \citenamefont {Mock}, \citenamefont {Korlacki}, \citenamefont {Knight},
  \citenamefont {Darakchieva}, \citenamefont {Schöche},\ and\ \citenamefont
  {Schubert}}]{StokeyJAP2020LSO}%
  \BibitemOpen
  \bibfield  {author} {\bibinfo {author} {\bibfnamefont {M.}~\bibnamefont
  {Stokey}}, \bibinfo {author} {\bibfnamefont {A.}~\bibnamefont {Mock}},
  \bibinfo {author} {\bibfnamefont {R.}~\bibnamefont {Korlacki}}, \bibinfo
  {author} {\bibfnamefont {S.}~\bibnamefont {Knight}}, \bibinfo {author}
  {\bibfnamefont {V.}~\bibnamefont {Darakchieva}}, \bibinfo {author}
  {\bibfnamefont {S.}~\bibnamefont {Schöche}}, \ and\ \bibinfo {author}
  {\bibfnamefont {M.}~\bibnamefont {Schubert}},\ }\bibfield  {title} {\enquote
  {\bibinfo {title} {Infrared active phonons in monoclinic lutetium
  oxyorthosilicate},}\ }\href {\doibase 10.1063/1.5135016} {\bibfield
  {journal} {\bibinfo  {journal} {Journal of Applied Physics}\ }\textbf
  {\bibinfo {volume} {127}},\ \bibinfo {pages} {115702} (\bibinfo {year}
  {2020})},\ \Eprint {http://arxiv.org/abs/https://doi.org/10.1063/1.5135016}
  {https://doi.org/10.1063/1.5135016} \BibitemShut {NoStop}%
\bibitem [{\citenamefont {Schubert}(2016)}]{SchubertPRL2016GLST}%
  \BibitemOpen
  \bibfield  {author} {\bibinfo {author} {\bibfnamefont {M.}~\bibnamefont
  {Schubert}},\ }\bibfield  {title} {\enquote {\bibinfo {title}
  {Coordinate-invariant lyddane-sachs-teller relationship for polar vibrations
  in materials with monoclinic and triclinic crystal systems},}\ }\href@noop {}
  {\bibfield  {journal} {\bibinfo  {journal} {Phys. Rev. Lett.}\ }\textbf
  {\bibinfo {volume} {117}},\ \bibinfo {pages} {215502} (\bibinfo {year}
  {2016})}\BibitemShut {NoStop}%
\bibitem [{\citenamefont {Yiwen}, \citenamefont {Yao},\ and\ \citenamefont
  {Wang}(2018)}]{yiwen2018propagation}%
  \BibitemOpen
  \bibfield  {author} {\bibinfo {author} {\bibfnamefont {E.}~\bibnamefont
  {Yiwen}}, \bibinfo {author} {\bibfnamefont {J.}~\bibnamefont {Yao}}, \ and\
  \bibinfo {author} {\bibfnamefont {L.}~\bibnamefont {Wang}},\ }\bibfield
  {title} {\enquote {\bibinfo {title} {Propagation of terahertz waves in a
  monoclinic crystal \uppercase{B}a\uppercase{G}a$_4$\uppercase{S}e$_7$},}\
  }\href@noop {} {\bibfield  {journal} {\bibinfo  {journal} {Scientific
  reports}\ }\textbf {\bibinfo {volume} {8}},\ \bibinfo {pages} {1--8}
  (\bibinfo {year} {2018})}\BibitemShut {NoStop}%
\bibitem [{\citenamefont {Ghosh}\ and\ \citenamefont
  {Singisetti}(2016)}]{Ghosh2016_Ab_initio}%
  \BibitemOpen
  \bibfield  {author} {\bibinfo {author} {\bibfnamefont {K.}~\bibnamefont
  {Ghosh}}\ and\ \bibinfo {author} {\bibfnamefont {U.}~\bibnamefont
  {Singisetti}},\ }\bibfield  {title} {\enquote {\bibinfo {title} {Ab initio
  calculation of electron–phonon coupling in monoclinic
  $\beta$-\uppercase{G}a$_2$\uppercase{O}$_3$ crystal},}\ }\href@noop {}
  {\bibfield  {journal} {\bibinfo  {journal} {Applied Physics Letters}\
  }\textbf {\bibinfo {volume} {109}},\ \bibinfo {pages} {072102} (\bibinfo
  {year} {2016})}\BibitemShut {NoStop}%
\bibitem [{\citenamefont {Kang}\ \emph {et~al.}(2017)\citenamefont {Kang},
  \citenamefont {Krishnaswamy}, \citenamefont {Peelaers},\ and\ \citenamefont
  {Walle}}]{Kang2017_Fund_limits}%
  \BibitemOpen
  \bibfield  {author} {\bibinfo {author} {\bibfnamefont {Y.}~\bibnamefont
  {Kang}}, \bibinfo {author} {\bibfnamefont {K.}~\bibnamefont {Krishnaswamy}},
  \bibinfo {author} {\bibfnamefont {H.}~\bibnamefont {Peelaers}}, \ and\
  \bibinfo {author} {\bibfnamefont {C.~G. V.~d.}\ \bibnamefont {Walle}},\
  }\bibfield  {title} {\enquote {\bibinfo {title} {Fundamental limits on the
  electron mobility of $\beta$-\uppercase{G}a$_2$\uppercase{O}$_3$},}\
  }\href@noop {} {\bibfield  {journal} {\bibinfo  {journal} {Journal of
  Physics: Condensed Matter}\ }\textbf {\bibinfo {volume} {29}},\ \bibinfo
  {pages} {234001} (\bibinfo {year} {2017})}\BibitemShut {NoStop}%
\bibitem [{\citenamefont {Geller}(1960)}]{GellerJPC1960Ga2O3}%
  \BibitemOpen
  \bibfield  {author} {\bibinfo {author} {\bibfnamefont {S.}~\bibnamefont
  {Geller}},\ }\bibfield  {title} {\enquote {\bibinfo {title} {Crystal
  structure of $\beta$-\uppercase{G}a$_2$\uppercase{O}$_3$},}\ }\href@noop {}
  {\bibfield  {journal} {\bibinfo  {journal} {J. Chem. Phys.}\ }\textbf
  {\bibinfo {volume} {33}},\ \bibinfo {pages} {676} (\bibinfo {year}
  {1960})}\BibitemShut {NoStop}%
\bibitem [{\citenamefont {Mueller}(1943)}]{MuellerMIT1943}%
  \BibitemOpen
  \bibfield  {author} {\bibinfo {author} {\bibfnamefont {H.}~\bibnamefont
  {Mueller}},\ }\href@noop {} {\enquote {\bibinfo {title} {Memorandum on the
  polarization optics of the photoelastic shutter},}\ }\bibinfo {type} {Report
  of the OSRD project OEMsr-576}\ \bibinfo {number} {2}\ (\bibinfo
  {institution} {Massachusets Institute of Technology},\ \bibinfo {year}
  {1943})\BibitemShut {NoStop}%
\bibitem [{\citenamefont
  {Schubert}(2004{\natexlab{a}})}]{SchubertIRSEBook_2004}%
  \BibitemOpen
  \bibfield  {author} {\bibinfo {author} {\bibfnamefont {M.}~\bibnamefont
  {Schubert}},\ }\href@noop {} {\emph {\bibinfo {title} {Infrared Ellipsometry
  on semiconductor layer structures: Phonons, plasmons and polaritons}}},\
  \bibinfo {series} {Springer Tracts in Modern Physics}, Vol.\ \bibinfo
  {volume} {209}\ (\bibinfo  {publisher} {Springer},\ \bibinfo {address}
  {Berlin},\ \bibinfo {year} {2004})\BibitemShut {NoStop}%
\bibitem [{\citenamefont {Fujiwara}(2007)}]{Fujiwara_2007}%
  \BibitemOpen
  \bibfield  {author} {\bibinfo {author} {\bibfnamefont {H.}~\bibnamefont
  {Fujiwara}},\ }\href@noop {} {\emph {\bibinfo {title} {Spectroscopic
  Ellipsometry}}}\ (\bibinfo  {publisher} {John Wiley \& Sons},\ \bibinfo
  {address} {New York},\ \bibinfo {year} {2007})\BibitemShut {NoStop}%
\bibitem [{\citenamefont {K\"{u}hne}\ \emph {et~al.}(2014)\citenamefont
  {K\"{u}hne}, \citenamefont {Herzinger}, \citenamefont {Schubert},
  \citenamefont {Woollam},\ and\ \citenamefont {Hofmann}}]{KuehneRSI_2014}%
  \BibitemOpen
  \bibfield  {author} {\bibinfo {author} {\bibfnamefont {P.}~\bibnamefont
  {K\"{u}hne}}, \bibinfo {author} {\bibfnamefont {C.~M.}\ \bibnamefont
  {Herzinger}}, \bibinfo {author} {\bibfnamefont {M.}~\bibnamefont {Schubert}},
  \bibinfo {author} {\bibfnamefont {J.~A.}\ \bibnamefont {Woollam}}, \ and\
  \bibinfo {author} {\bibfnamefont {T.}~\bibnamefont {Hofmann}},\ }\bibfield
  {title} {\enquote {\bibinfo {title} {An integrated mid-infrared, far-infrared
  and terahertz optical hall effect instrument},}\ }\href {\doibase xxx}
  {\bibfield  {journal} {\bibinfo  {journal} {Rev. Sci. Instrum.}\ }\textbf
  {\bibinfo {volume} {85}},\ \bibinfo {pages} {071301} (\bibinfo {year}
  {2014})}\BibitemShut {NoStop}%
\bibitem [{\citenamefont {Schubert}(2004{\natexlab{b}})}]{Schubert04}%
  \BibitemOpen
  \bibfield  {author} {\bibinfo {author} {\bibfnamefont {M.}~\bibnamefont
  {Schubert}},\ }\bibfield  {title} {\enquote {\bibinfo {title} {Theory and
  application of generalized ellipsometry},}\ }in\ \href@noop {} {\emph
  {\bibinfo {booktitle} {Handbook of Ellipsometry}}},\ \bibinfo {editor}
  {edited by\ \bibinfo {editor} {\bibfnamefont {E.}~\bibnamefont {Irene}}\ and\
  \bibinfo {editor} {\bibfnamefont {H.}~\bibnamefont {Tompkins}}}\ (\bibinfo
  {publisher} {William Andrew Publishing},\ \bibinfo {year} {2004})\BibitemShut
  {NoStop}%
\bibitem [{\citenamefont {Galazka}\ \emph {et~al.}(2018)\citenamefont
  {Galazka}, \citenamefont {Ganschow}, \citenamefont {Fiedler}, \citenamefont
  {Bertram}, \citenamefont {Klimm}, \citenamefont {Irmscher}, \citenamefont
  {Schewski}, \citenamefont {Pietsch}, \citenamefont {Albrecht},\ and\
  \citenamefont {Bickermann}}]{GALAZKA201882}%
  \BibitemOpen
  \bibfield  {author} {\bibinfo {author} {\bibfnamefont {Z.}~\bibnamefont
  {Galazka}}, \bibinfo {author} {\bibfnamefont {S.}~\bibnamefont {Ganschow}},
  \bibinfo {author} {\bibfnamefont {A.}~\bibnamefont {Fiedler}}, \bibinfo
  {author} {\bibfnamefont {R.}~\bibnamefont {Bertram}}, \bibinfo {author}
  {\bibfnamefont {D.}~\bibnamefont {Klimm}}, \bibinfo {author} {\bibfnamefont
  {K.}~\bibnamefont {Irmscher}}, \bibinfo {author} {\bibfnamefont
  {R.}~\bibnamefont {Schewski}}, \bibinfo {author} {\bibfnamefont
  {M.}~\bibnamefont {Pietsch}}, \bibinfo {author} {\bibfnamefont
  {M.}~\bibnamefont {Albrecht}}, \ and\ \bibinfo {author} {\bibfnamefont
  {M.}~\bibnamefont {Bickermann}},\ }\bibfield  {title} {\enquote {\bibinfo
  {title} {Doping of czochralski-grown bulk
  $\beta$-\uppercase{G}a$_2$\uppercase{O}$_3$ single crystals with cr, ce and
  al},}\ }\href {\doibase https://doi.org/10.1016/j.jcrysgro.2018.01.022}
  {\bibfield  {journal} {\bibinfo  {journal} {Journal of Crystal Growth}\
  }\textbf {\bibinfo {volume} {486}},\ \bibinfo {pages} {82 -- 90} (\bibinfo
  {year} {2018})}\BibitemShut {NoStop}%
\bibitem [{\citenamefont {Galazka}\ \emph {et~al.}(2020)\citenamefont
  {Galazka}, \citenamefont {Irmscher}, \citenamefont {Schewski}, \citenamefont
  {Hanke}, \citenamefont {Pietsch}, \citenamefont {Ganschow}, \citenamefont
  {Klimm}, \citenamefont {Dittmar}, \citenamefont {Fiedler}, \citenamefont
  {Schroeder},\ and\ \citenamefont {Bickermann}}]{GALAZKA2020125297}%
  \BibitemOpen
  \bibfield  {author} {\bibinfo {author} {\bibfnamefont {Z.}~\bibnamefont
  {Galazka}}, \bibinfo {author} {\bibfnamefont {K.}~\bibnamefont {Irmscher}},
  \bibinfo {author} {\bibfnamefont {R.}~\bibnamefont {Schewski}}, \bibinfo
  {author} {\bibfnamefont {I.~M.}\ \bibnamefont {Hanke}}, \bibinfo {author}
  {\bibfnamefont {M.}~\bibnamefont {Pietsch}}, \bibinfo {author} {\bibfnamefont
  {S.}~\bibnamefont {Ganschow}}, \bibinfo {author} {\bibfnamefont
  {D.}~\bibnamefont {Klimm}}, \bibinfo {author} {\bibfnamefont
  {A.}~\bibnamefont {Dittmar}}, \bibinfo {author} {\bibfnamefont
  {A.}~\bibnamefont {Fiedler}}, \bibinfo {author} {\bibfnamefont
  {T.}~\bibnamefont {Schroeder}}, \ and\ \bibinfo {author} {\bibfnamefont
  {M.}~\bibnamefont {Bickermann}},\ }\bibfield  {title} {\enquote {\bibinfo
  {title} {Czochralski-grown bulk \uppercase{G}a$_2$\uppercase{O}$_3$ single
  crystals doped with mono-, di-, tri-, and tetravalent ions},}\ }\href
  {\doibase https://doi.org/10.1016/j.jcrysgro.2019.125297} {\bibfield
  {journal} {\bibinfo  {journal} {Journal of Crystal Growth}\ }\textbf
  {\bibinfo {volume} {529}},\ \bibinfo {pages} {125297} (\bibinfo {year}
  {2020})}\BibitemShut {NoStop}%
\bibitem [{\citenamefont {Liu}, \citenamefont {Gu},\ and\ \citenamefont
  {Liu}(2007)}]{liu2007lattice}%
  \BibitemOpen
  \bibfield  {author} {\bibinfo {author} {\bibfnamefont {B.}~\bibnamefont
  {Liu}}, \bibinfo {author} {\bibfnamefont {M.}~\bibnamefont {Gu}}, \ and\
  \bibinfo {author} {\bibfnamefont {X.}~\bibnamefont {Liu}},\ }\bibfield
  {title} {\enquote {\bibinfo {title} {Lattice dynamical, dielectric, and
  thermodynamic properties of $\beta$-\uppercase{G}a$_2$\uppercase{O}$_3$ from
  first principles},}\ }\href@noop {} {\bibfield  {journal} {\bibinfo
  {journal} {Applied Physics Letters}\ }\textbf {\bibinfo {volume} {91}},\
  \bibinfo {pages} {172102} (\bibinfo {year} {2007})}\BibitemShut {NoStop}%
\bibitem [{\citenamefont {Schubert}\ \emph
  {et~al.}(2019{\natexlab{b}})\citenamefont {Schubert}, \citenamefont {Mock},
  \citenamefont {Korlacki}, \citenamefont {Knight}, \citenamefont {Galazka},
  \citenamefont {Wagner}, \citenamefont {Wheeler}, \citenamefont {Tadjer},
  \citenamefont {Goto},\ and\ \citenamefont
  {Darakchieva}}]{schubert2019longitudinal}%
  \BibitemOpen
  \bibfield  {author} {\bibinfo {author} {\bibfnamefont {M.}~\bibnamefont
  {Schubert}}, \bibinfo {author} {\bibfnamefont {A.}~\bibnamefont {Mock}},
  \bibinfo {author} {\bibfnamefont {R.}~\bibnamefont {Korlacki}}, \bibinfo
  {author} {\bibfnamefont {S.}~\bibnamefont {Knight}}, \bibinfo {author}
  {\bibfnamefont {Z.}~\bibnamefont {Galazka}}, \bibinfo {author} {\bibfnamefont
  {G.}~\bibnamefont {Wagner}}, \bibinfo {author} {\bibfnamefont
  {V.}~\bibnamefont {Wheeler}}, \bibinfo {author} {\bibfnamefont
  {M.}~\bibnamefont {Tadjer}}, \bibinfo {author} {\bibfnamefont
  {K.}~\bibnamefont {Goto}}, \ and\ \bibinfo {author} {\bibfnamefont
  {V.}~\bibnamefont {Darakchieva}},\ }\bibfield  {title} {\enquote {\bibinfo
  {title} {Longitudinal phonon plasmon mode coupling in
  $\beta$-\uppercase{G}a$_2$\uppercase{O}$_3$},}\ }\href@noop {} {\bibfield
  {journal} {\bibinfo  {journal} {Applied Physics Letters}\ }\textbf {\bibinfo
  {volume} {114}},\ \bibinfo {pages} {102102} (\bibinfo {year}
  {2019}{\natexlab{b}})}\BibitemShut {NoStop}%
\bibitem [{\citenamefont {Matsumoto}\ \emph {et~al.}(2008)\citenamefont
  {Matsumoto}, \citenamefont {Hosokura}, \citenamefont {Kageyama},
  \citenamefont {Takagi}, \citenamefont {Sakabe},\ and\ \citenamefont
  {Hangyo}}]{matsumoto2008analysis}%
  \BibitemOpen
  \bibfield  {author} {\bibinfo {author} {\bibfnamefont {N.}~\bibnamefont
  {Matsumoto}}, \bibinfo {author} {\bibfnamefont {T.}~\bibnamefont {Hosokura}},
  \bibinfo {author} {\bibfnamefont {K.}~\bibnamefont {Kageyama}}, \bibinfo
  {author} {\bibfnamefont {H.}~\bibnamefont {Takagi}}, \bibinfo {author}
  {\bibfnamefont {Y.}~\bibnamefont {Sakabe}}, \ and\ \bibinfo {author}
  {\bibfnamefont {M.}~\bibnamefont {Hangyo}},\ }\bibfield  {title} {\enquote
  {\bibinfo {title} {Analysis of dielectric response of
  \uppercase{T}i\uppercase{O}$_2$ in terahertz frequency region by general
  harmonic oscillator model},}\ }\href@noop {} {\bibfield  {journal} {\bibinfo
  {journal} {Japanese journal of applied physics}\ }\textbf {\bibinfo {volume}
  {47}},\ \bibinfo {pages} {7725} (\bibinfo {year} {2008})}\BibitemShut
  {NoStop}%
\bibitem [{\citenamefont {J{\"o}rdens}\ \emph {et~al.}(2009)\citenamefont
  {J{\"o}rdens}, \citenamefont {Scheller}, \citenamefont {Wichmann},
  \citenamefont {Mikulics}, \citenamefont {Wiesauer},\ and\ \citenamefont
  {Koch}}]{jordens2009terahertz}%
  \BibitemOpen
  \bibfield  {author} {\bibinfo {author} {\bibfnamefont {C.}~\bibnamefont
  {J{\"o}rdens}}, \bibinfo {author} {\bibfnamefont {M.}~\bibnamefont
  {Scheller}}, \bibinfo {author} {\bibfnamefont {M.}~\bibnamefont {Wichmann}},
  \bibinfo {author} {\bibfnamefont {M.}~\bibnamefont {Mikulics}}, \bibinfo
  {author} {\bibfnamefont {K.}~\bibnamefont {Wiesauer}}, \ and\ \bibinfo
  {author} {\bibfnamefont {M.}~\bibnamefont {Koch}},\ }\bibfield  {title}
  {\enquote {\bibinfo {title} {Terahertz birefringence for orientation
  analysis},}\ }\href@noop {} {\bibfield  {journal} {\bibinfo  {journal}
  {Applied optics}\ }\textbf {\bibinfo {volume} {48}},\ \bibinfo {pages}
  {2037--2044} (\bibinfo {year} {2009})}\BibitemShut {NoStop}%
\bibitem [{\citenamefont {Matsumoto}\ \emph {et~al.}(2009)\citenamefont
  {Matsumoto}, \citenamefont {Fujii}, \citenamefont {Kageyama}, \citenamefont
  {Takagi}, \citenamefont {Nagashima},\ and\ \citenamefont
  {Hangyo}}]{matsumoto2009measurement}%
  \BibitemOpen
  \bibfield  {author} {\bibinfo {author} {\bibfnamefont {N.}~\bibnamefont
  {Matsumoto}}, \bibinfo {author} {\bibfnamefont {T.}~\bibnamefont {Fujii}},
  \bibinfo {author} {\bibfnamefont {K.}~\bibnamefont {Kageyama}}, \bibinfo
  {author} {\bibfnamefont {H.}~\bibnamefont {Takagi}}, \bibinfo {author}
  {\bibfnamefont {T.}~\bibnamefont {Nagashima}}, \ and\ \bibinfo {author}
  {\bibfnamefont {M.}~\bibnamefont {Hangyo}},\ }\bibfield  {title} {\enquote
  {\bibinfo {title} {Measurement of the soft-mode dispersion in
  \uppercase{S}r\uppercase{T}i\uppercase{O}$_3$ by terahertz time-domain
  spectroscopic ellipsometry},}\ }\href@noop {} {\bibfield  {journal} {\bibinfo
   {journal} {Japanese Journal of Applied Physics}\ }\textbf {\bibinfo {volume}
  {48}},\ \bibinfo {pages} {09KC11} (\bibinfo {year} {2009})}\BibitemShut
  {NoStop}%
\end{thebibliography}%
\end{document}


\begin{center}
\title{Supplementary Information}
\maketitle
\end{center}

\begin{flushleft}
\textit{A. Sample Information}
\end{flushleft}
\begin{ruledtabular}
\begin{tabular}{||c|c|c|c||}
Sample Name&Description&Orientation&Provider \\
\hline  
Sample A & \makecell{A single side polished (SSP) Czochralski grown  trivalent \\ (Al$^{3+}$) doped single crystal with 5~mol~\% of dopant \\ concentration in the melt and is electrically insulating}&(100) & IKZ\\
\hline
Sample B& \makecell{Czochralski grown single crystal that is Fe \\ compensated and electrically insulating with \\ a mechanically measured nominal thickness of 523$\mu$m}&(010)& \makecell{Northrop \\ Grumman \\ SYNOPTICS} \\
\hline
Sample C& \makecell{Fe compensated and electrically insulating single \\ crystal of mechanically measured thickness of 656 $\mu$m}&(001)& \makecell{Novel \\ Crystal \\ Technologies \\ Inc.}\\
\end{tabular}
\end{ruledtabular}

\begin{flushleft}
\textit{B. Analytical Models for CW transmission measurement}
\end{flushleft}

Given the multiple reflection scenario resulting from the ‘Air/GO’ and ‘GO/Air’ interfaces the ABCD matrix formalism was adopted to model the measured Fabry-Perot oscillations:

\begin{equation}\label{eq:S1}
\begin{pmatrix}
A & B\\
C & D\\                    
\end{pmatrix}
= 
\begin{pmatrix}
cos(\phi_{GO}) &  jZ_{GO}sin(\phi_{GO})\\
\frac{j}{Z_{GO}}sin(\phi_{GO}) & cos(\phi_{GO})\\    
\end{pmatrix}
\end{equation}

where $\phi_{GO} =\frac{{n}_{GO}\omega d_{GO}}{c}$ provides the propagation phase of the terahertz wave along a certain direction, ${n}_{GO}$ is the refractive index along a given direction, $\omega$ is the angular frequency, c is the speed of light, $d_{GO}$ is the sample thickness and,  $Z_{GO}=  \frac{Z_o}{{n}_{GO}}$  where $Z_o$ is the free-space impedance ($Z_o$ = 377 $\Omega$). 

The following formula represents the final (complex) terahertz transmission   through the substrate

\begin{equation}\label{eq:S2}
     t_{GO}= \frac{2}{A + \frac{B}{Z_o} +CZ_o+D}
\end{equation}

Equations (1) and (2) are employed for calculating transmission along the crystal axes. The transmission at other angles $\psi$ in the a-c plane was modeled as a superposition of the transmission response of the two (principal) axes formed by the eigenvectors. 
\begin{equation}\label{eq: S3}
t= t_1 cos^{2}(\psi)+ t_2sin^{2}(\psi)                 
\end{equation}
where '$\psi$' is the in-plane angle between the E-field polarization and the sample’s crystal axes.

\begin{flushleft}
\textit{C. Mueller matrix data of (010) Fe doped substrate} \\
\end{flushleft}
\begin{figure*}[!htbp]
        \includegraphics[scale=0.5, keepaspectratio = true]{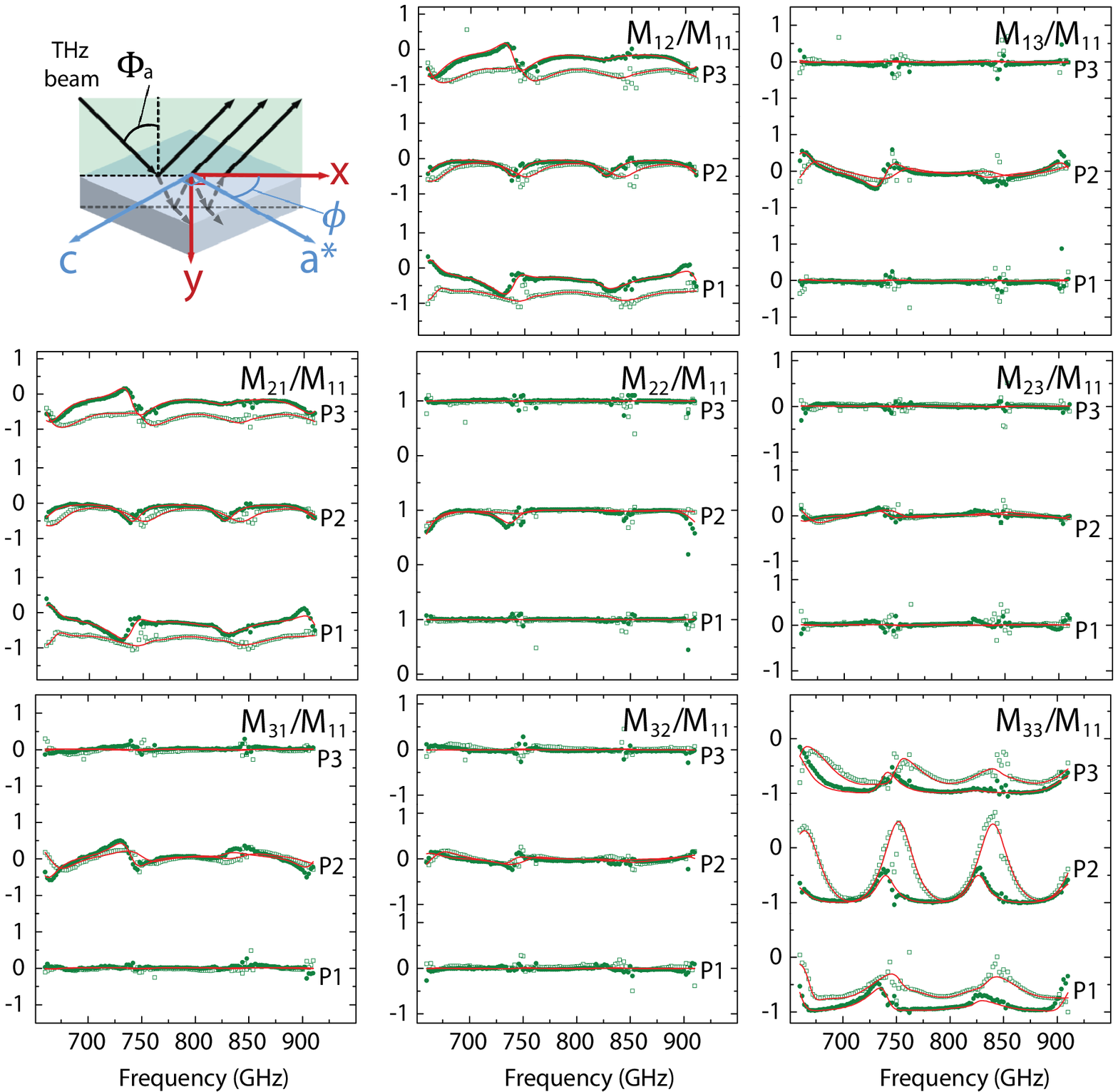}
    \caption{\label{fig:GSE} $\beta$-Ga$_2$O$_3$ (010) GSE data at $\Phi_a=40^{\circ}$ and $60^{\circ}$ angle of incidence: Green symbols (experiment); Solid red lines (best match model calculated). Solid green circles indicate data for $\Phi_a=40^{\circ}$, and open squares for $\Phi_a=60^{\circ}$. Data are presented in the Mueller matrix formalism. All data are normalized to element M$_{11}$. Data are shown for three azimuths: P1 [$\varphi=0.9\pm0.2^{\circ}$]; P2 [$\varphi=45.9\pm0.2^{\circ}$]; P3 [$\varphi=90.9\pm0.2^{\circ}$]. The inset depicts schematically the location of axes $\mathbf{a}^{\star}$ and $\mathbf{c}$ in position P2. Axes $x$ and $y$ indicate the ellipsometer coordinate axes, where $x$ is parallel to the plane of incidence. The sample thickness was determined from the same model analysis as $d=525\pm1\mu m$.}
    \label{THz_MM_(010)}
\end{figure*}

Figure~\ref{THz_MM_(010)} shows the THz-GSE Mueller matrix data for sample A (010) cut $\beta$-Ga$_2$O$_3$  for multiple angles of incidence and multiple azimuth orientations. This data was analyzed simultaneously with data shown in Fig. 2 (main manuscript) to produce the best-match model results for permittivity given in Table 1. In addition to real-valued contributions to the permittivity, a small imaginary contribution Im\{$\varepsilon$\} is also included in our model analysis to perfectly fit the experimental data (for both 010 and 100 samples). The values for Im\{$\varepsilon$\} which correspond to the $\mathbf{a}^{\star}$, $\mathbf{b}$, and $\mathbf{c}$ axes are 0.10$\pm$0.004, 0.08$\pm$0.004, and 0.11$\pm$0.004, respectively.

\begin{flushleft}
\textit{D. THz transmission in the a-c plane} \\
\end{flushleft}
\begin{figure*}[!htbp]
    \centering
    \includegraphics[scale=0.9, keepaspectratio = true]{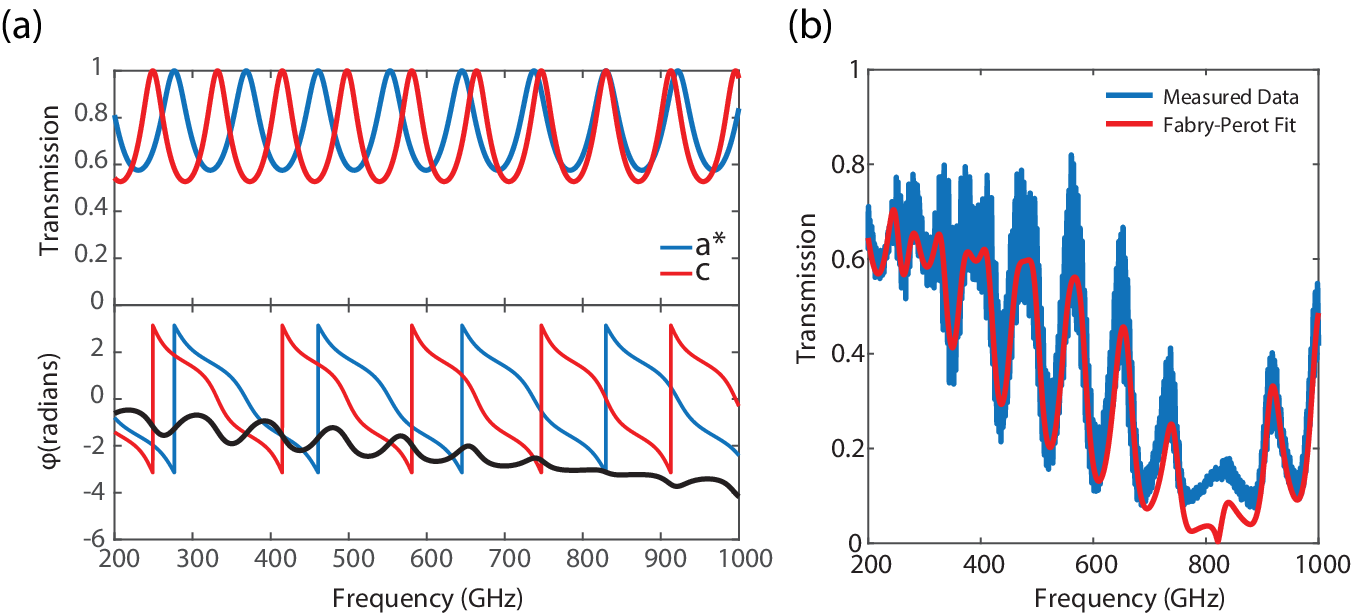}
    \caption{(a) (top) Calculated transmission amplitudes for polarization parallel to $\sim a^*$ and $\sim c$, (bottom) their corresponding phase difference illustrating the fact that phase difference becomes $\sim \pi rad $ at a specific frequency ($\sim 810$ GHz), (b) measured transmission at $\psi$ = $45^{0}$ in blue and the calculated response in orange}
    \label{45_deg_Response}
\end{figure*}
Another notable aspect is that in the a-c plane at intermediate angles between the principal dielectric axes, the measured transmission spectrum deviates from the single mode FPs shown in Fig.\ref{45_deg_Response}. As a result, at $\psi= 45^{o}$ significant polarization rotation can occur. Utilizing the extracted principal indices, the amplitude and phase of transmission were calculated for incident fields parallel to the two major polarizability axes $a^{*}$ and $c$ and are shown in the upper and lower panels of Fig. \ref{45_deg_Response}(a) respectively. We notice that the phase difference $(\Delta \phi)$ between the two becomes $180^{\circ}$ at a specific frequency point (and its multiples) . This phase difference between the two principal axes indicates that the transmitted polarization is rotated to the orthogonal direction of the input/detection polarization resulting in a transmission minima as seen in Fig. \ref{45_deg_Response}(b). The gallium oxide substrate (for the given thickness and birefringence) behaves as a half wave-plate. Consequently, the position of this minima is sensitive to the index difference $(\Delta n)$ of the constituent polarizability axes and the substrate thickness. We can see a good agreement between the measured and modelled response. 

Additionally, in Figure 3 (of the main manuscript) we observe that while at low end of the spectrum, i.e. frequency < 0.7 THz, the FP maxima reaches unity, as expected from the insulating nature of the samples, at the higher frequency end (> 0.7 THz), the maxima slightly drop from unity. This can be understood as arising from an apparent frequency dependent loss component, which can be represented through an effective conductivity ranging from ~0 S/m (at frequencies up to ~0.7 THz) to ~2 S/m (at ~1 THz). This small loss, which does not yield discernible index dispersion, does not stem from free-carrier concentration but could possibly be due to either i) the tail of phonon modes occurring at higher frequencies, ii) a manifestation of a small offset between the measured directions and the actual polarizability axes, or iii) a non-uniform sample thickness over the probe beam spot. Exploring the underlying reason for this observation requires further measurements, e.g. in an extended spectral range, and is beyond the scope the of the current study.